\documentclass[twocolumn,showpacs,preprintnumbers,amsmath,amssymb]{revtex4}
\usepackage{graphicx}% Include figure files
\usepackage{dcolumn}% Align table columns on decimal point
\usepackage{bm}% bold math
\usepackage{color}
\begin{document}

\title{%Soliton- and shock-induced 
Negative-frequency dispersive wave generation in quadratic media}
%\title{Generalized multiple dispersive wave generation in quadratic media}

\author{Matteo Conforti$^1$, Niclas Westerberg$^2$, Fabio Baronio$^1$, Stefano Trillo$^3$, and Daniele Faccio$^2$}

\affiliation{$^1$CNISM, Dipartimento di Ingegneria dell'Informazione, Universit\`a di Brescia, Via Branze 38, 25123 Brescia, Italy\\
$^2$ School of Engineering and Physical Sciences, SUPA, Heriot-Watt University, Edinburgh EH14 4AS, United Kingdom\\ 
$^3$ Dipartimento di Ingegneria, Universit\`a di Ferrara, Via Saragat 1, 44122 Ferrara, Italy}
\date{\today}

\begin{abstract}
We show that the extremely blue-shifted dispersive wave emitted in Kerr media owing to the coupling with the negative-frequency branch [Phys. Rev. Lett. {\bf 108}, 253901 (2012)] can be observed in quadratic media via second-harmonic generation. Not only such phenomenon is thus independent on the specific nonlinear mechanism, but it is shown to occur regardless of the fact that the process is pumped by a pulse which exhibits soliton-like features or, viceversa, undergoes wave-breaking. A simple unified formula gives the frequencies of the emitted dispersive waves in both cases.
%We give a simple recipe for calculating the frequencies of the emitted dispersive waves in both cases. 
%We show here that the process of negative-frequency dispersive wave generation is very general, and it is not restricted to a particular kind of media, nor to a particular source. In fact, conversely to what it is commonly believed, it is not strictly related to solitons,  and can be initiated either by solitary waves or by dispersive shock waves. Simulations in different quadratic media confirm our predictions.
\end{abstract}

\maketitle

\emph{Introduction}. Solitons emit resonant radiation (RR) 
%(also termed dispersive wave or Cherenkov radiation), 
owing to a universal mechanism of phase-matching with linear waves ruled by perturbing higher-order dispersive terms. Well-known examples range from fiber \cite{Wai,AK95} to Langmuir plasma  \cite{Karpman97} or water wave \cite{Karpman98} solitons. During the last decade, optical fibers offered the unique opportunity to deepen the physics of RR \cite{Skryabin03,BSY04,SG07}, with important applicative fall-out in supercontinuum generation \cite{DudleyRMP06}, where RR is responsible for broadening the spectrum over the blue-shifted (normally dispersive) region \cite{Cristiani04}. 
More recently, the field was significantly advanced by important results recognizing the role of RR in turbulence transport \cite{Rumpf09},  the observation of RR in different settings encompassing tapered \cite{Stark11} and noble-gas-filled photonic crystal fibers \cite{Joly11,Saleh11}, slow-light waveguides \cite{Colman12}, spatial diffraction in arrays \cite{Tran12}, and second-harmonic generation (SHG) \cite{Bache10,Zhou12}. 
Importantly, it was also shown that, in Kerr media, new frequencies  can be generated owing to the coupling with the negative-frequency part of the spectrum, a process termed negative-frequency resonant radiation (NRR) \cite{rubino12,rubino12b}. With reference to this new phenomenon, the aim of this letter is twofold: (i) to assess the universal nature of NRR by showing that it can be predicted to occur also via pure quadratic nonlinearities, under experimentally viable conditions of SHG; (ii) to generalize the concept of radiation by showing that in fact one does not necessarily need a soliton-like excitation, since RR and NRR can be produced also in the {\em opposite regime} where the nonlinearity, instead of compensating the action of group-velocity dispersion (GVD), is such to strongly enforce it leading to wave-breaking (shock formation) \cite{Rothenberg89}. This regime investigated recently in experiments performed in the spatial domain \cite{Hoefer06,Wan07} features dispersive shock wave (DSW), as predicted in a seminal work by Gurevich and Pitaevskii \cite{GP74}.

{\it Resonant radiation.} Let us first explain the general origin of the NRR  from a different perspective as compared with the analysis of Ref. \cite{rubino12,rubino12b}.
We consider an intense pump at carrier frequency $\omega_p$, characterized by a complex envelope $e(t,z)$, which travels with characteristic group-velocity $v$.
When such a pulse travels in a nonlinear medium without experiencing significant dispersive effects, its total electric field, which is by definition a real quantity, can be written as $E_p(z,\tau)=e(\tau) \exp[i  \overline k(\omega_p) z - i \omega_p \tau] + e^*(\tau) \exp[i  \overline k(-\omega_p) z + i \omega_p \tau]$, where $\tau=t-z/v$ is the retarded time and $ \overline k(\omega_p)=k(\omega_p)-\omega_p/v + k_{NL}(\omega_p)$, where $k(\omega_p)-\omega_p/v$ is the linear wave-number in the moving frame, $k_{NL}$ is the nonlinear correction (phase-shift) due to the nonlinearity, and $\overline k(-\omega_p)=-\overline k(\omega_p)$ for the field to be real.
Also linear waves (radiation) at frequency $\omega$ can be expressed in terms of positive- and negative-frequency content through the real field $E_r(z,t)=A(z) \exp[i k(\omega) z - i \omega t] + A^*(z) \exp[i k(-\omega) z + i \omega t]$. Here $k(\omega)=\omega n(\omega)/c$ is determined by the full dispersive relationship of the material in terms of the real index $n(\omega)=n(-\omega)$, neglecting losses for simplicity. Upon substitution $t=\tau + z/v$, we cast the radiation in the form $E_r(z,\tau)=A(z) \exp[i D(\omega) z - i \omega \tau] + A^*(z) \exp[-i D(\omega) z + i \omega \tau]$, where $D(\omega)=k(\omega)-\omega/v$ stands for the wave-number in a frame co-moving with the pump envelope $e$. Focusing on the positive frequency content of the radiation, its amplitude $A(z)$ starts to grow due to resonant transfer of energy from the pump at the phase-matching frequency $\omega=\omega_{RR}$ such that $\overline k(\omega_p)=D(\omega_{RR})$, which gives the well-known condition for RR \cite{Wai,AK95,SG07,Colman12,Bache10}. Conversely, what it is usually not recognized is the fact that $A(z)$ can be phase-matched also with the negative-frequency content of the pump at a different frequency $\omega=\omega_{NRR}$ such that the condition $-\overline k(\omega_p)=D(\omega_{NRR})$ is fulfilled.
By summarizing, both the phase-matched frequencies $\omega=\omega_{RR}, \omega_{NRR}$ can be obtained by solving a unified set of two equations
\begin{equation}\label{eqRR}
D(\omega)=\pm \overline k(\omega_p).
\end{equation}
We point out that, owing to symmetry, Eqs. (\ref{eqRR}) have also solutions $\omega=-\omega_{RR}, -\omega_{NRR}$, i.e. the image frequencies required by the Hermitian symmetry of the radiation field. Moreover, we arrive at Eq. (\ref{eqRR}) also when starting from the negative frequency content of the radiation $A^*$.

{\it Quadratic media.} In the following, we address the open question as to whether NRR can be observed in a quadratic medium. Our aim is to show that, in such media, Eqs. (\ref{eqRR}) accurately predict both the RR and NRR frequencies even if the pump pulse is not strictly invariant (strictly non-dispersive), provided one is able to accurately estimate its velocity $v$. In fact, a deviation from the ideal non-dispersive behavior of the pump is even beneficial since the growth of RR and NRR become significant when the pump undergoes a strong spectral broadening thereby seeding the phase-matched radiation frequencies. When nonlinearity and GVD act so as to mutually balance each other, this requires to operate with pulses which exhibit compression, i.e. higher-order solitons. However we go further by showing that also in the opposite regime, where a pulse experience strong temporal broadening, RR and NRR generation of comparable magnitude can be emitted. In order to demonstrate this, since the radiation detunings can be extremely large, we employ a description based on numerical integration of the $\chi^{(2)}$ Unidirectional Pulse Propagation Equation (UPPE2) implemented for anisotropic media \cite{Conforti11}. The latter is suitable to describe ultra-broadband propagation, not relying on the separation of spectral envelopes around the carriers \cite{Bache10,reviewSHG}), whose validity breaks down in the regime considered here. Furthermore this approach allows us to account for the full (all orders) dispersion $n(\omega)$ through the Sellmeier equations which characterize any specific material \cite{nikogosyan}.
%--fig 1
\begin{figure}[tb]
\includegraphics[width=8cm]{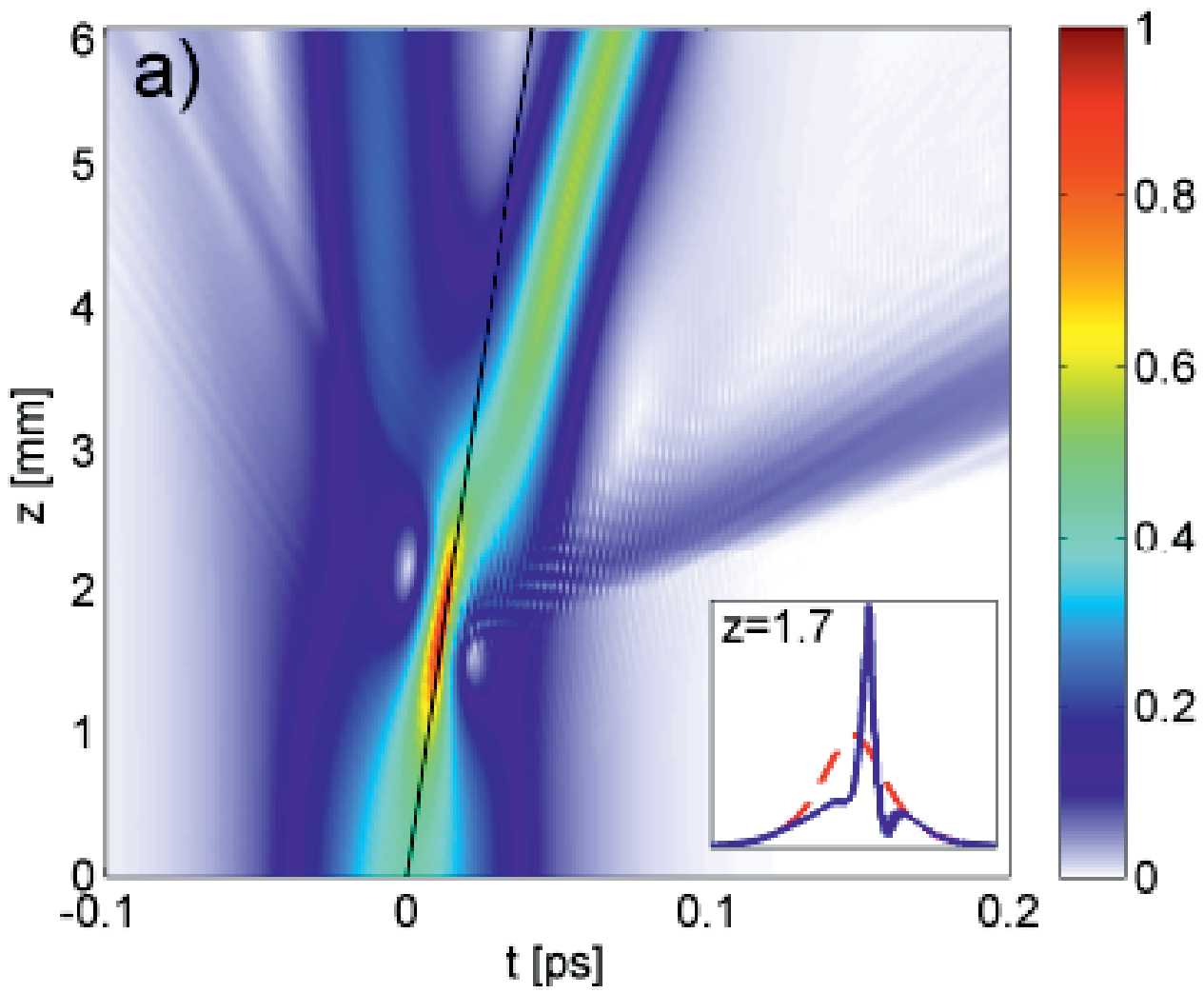}
\includegraphics[width=8cm]{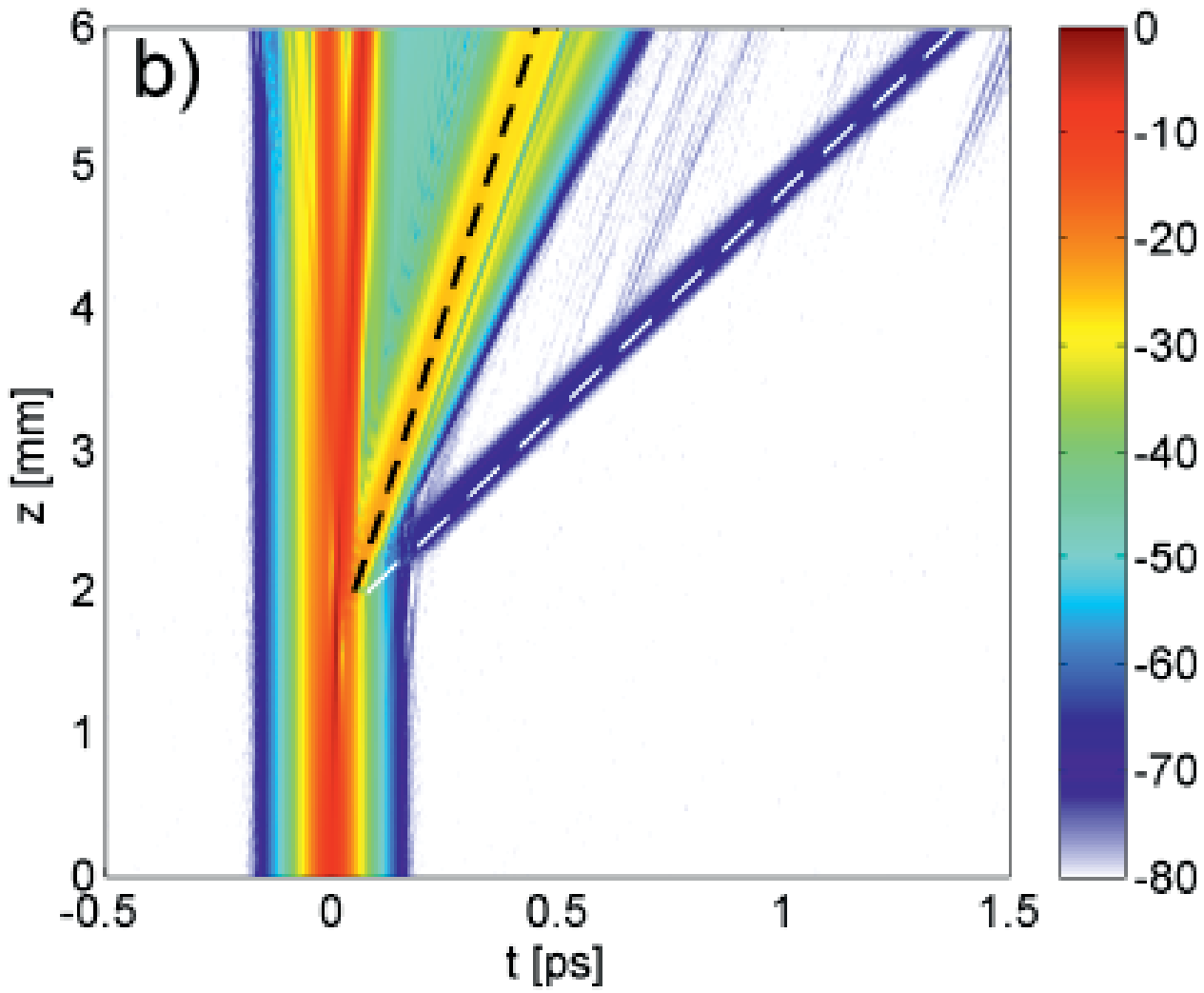}
\caption{(Color online) Time domain electric field \cite{nota1} evolution in a BBO crystal (ordinary polarization). (a) Linear scale.  The black line is the trajectory of a pulse propagating with group velocity $v_g=0.99876\cdot v_g(\omega_0)$. Inset: snapshot at the point maximal compression compared with the input. (b) Same in log scale (dB). Here the dashed black and dashed white lines follow the peak of the RR ($\lambda=405$ nm) and the NRR  ($\lambda=575$ nm), respectively. 
%Hyperbolic secant input (sech$(t/T_0)$) with peak intensity $I=7.4$ TW/cm$^2$.
} 
\label{fig1}
\end{figure}
%--fig 2
\begin{figure}[tb]
\includegraphics[width=8cm]{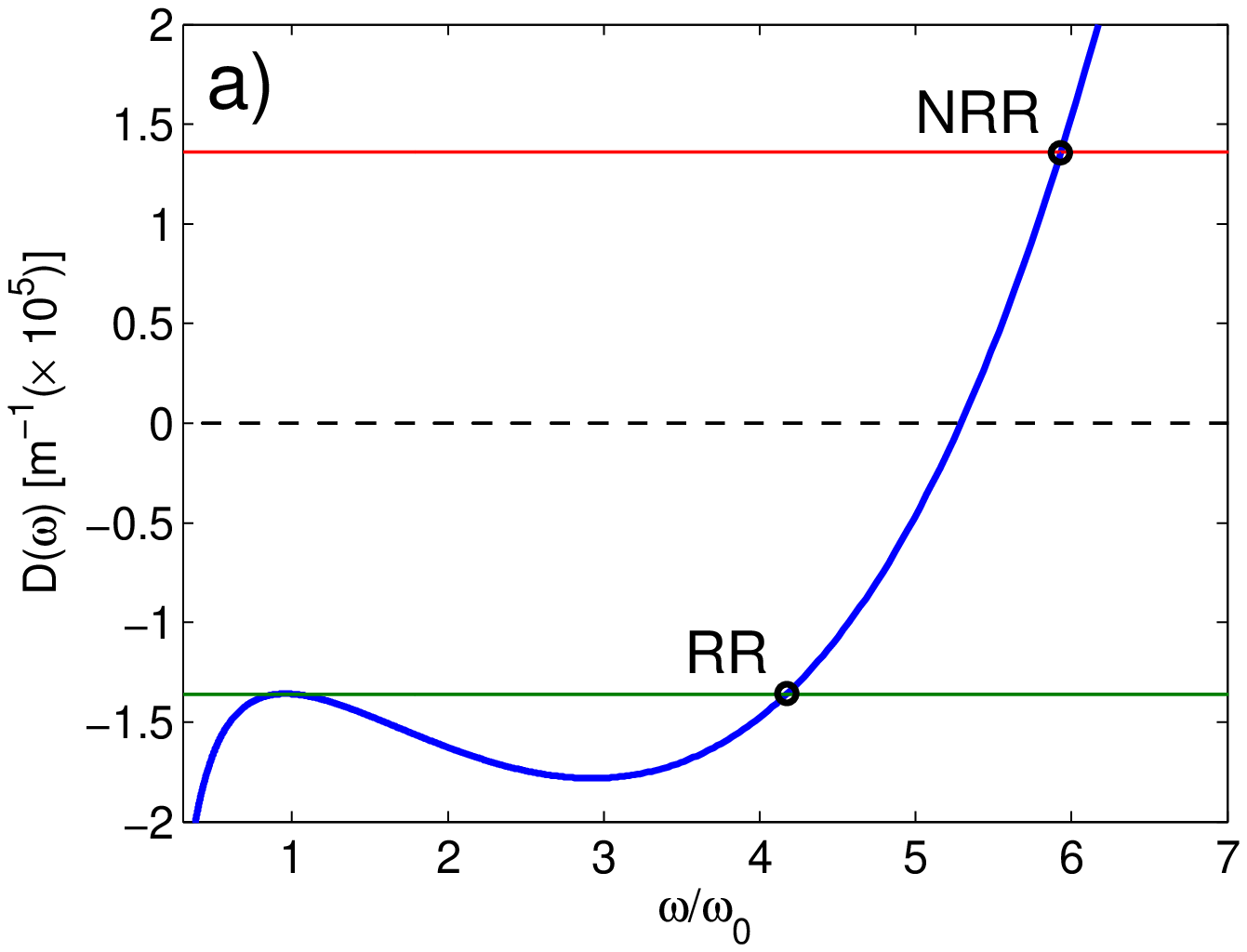}
\includegraphics[width=8cm]{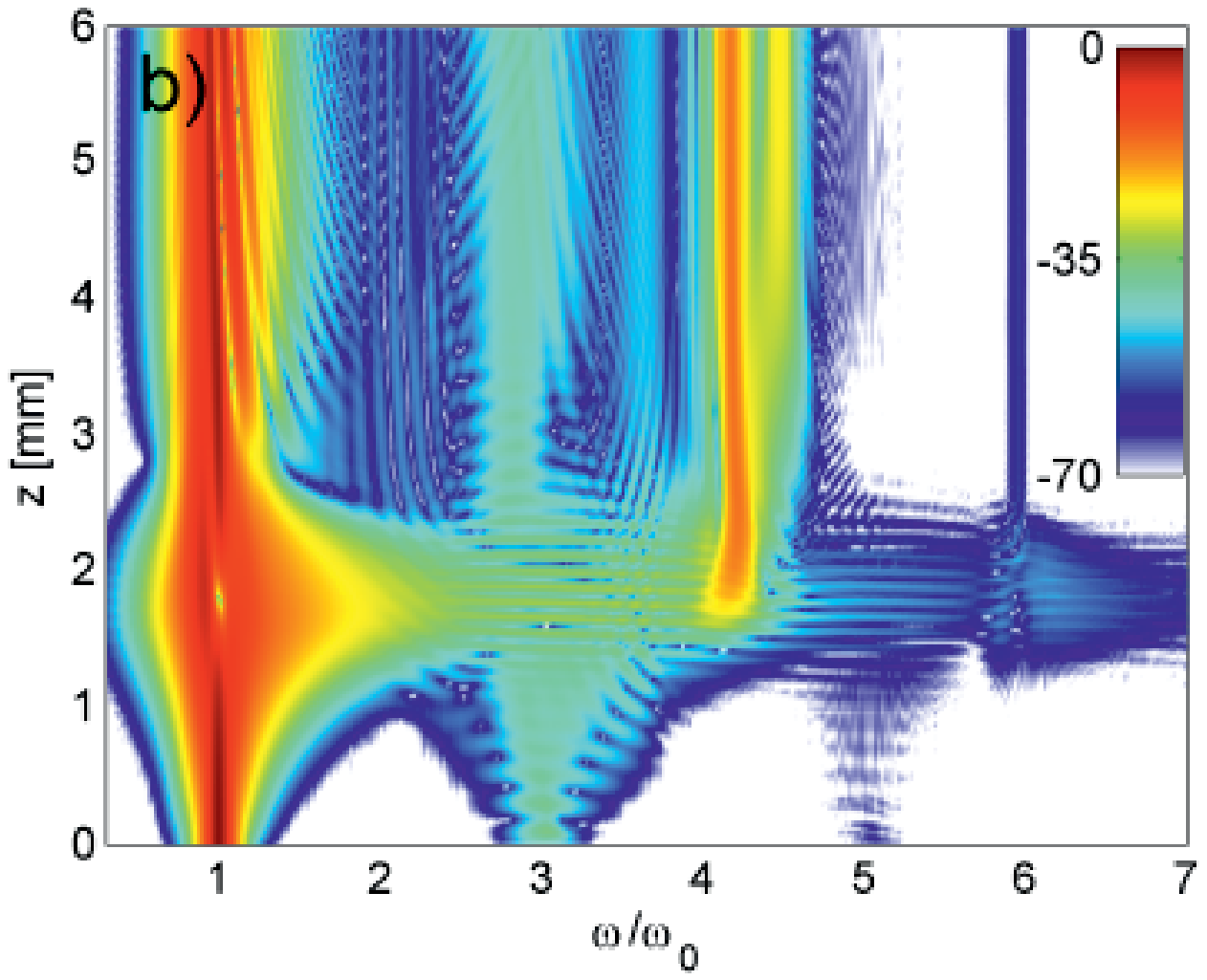}
\caption{(Color online) (a) Graphical solution of Eq.(\ref{eqRR}):  the intersections between the blue curve $D(\omega)$ and the red and green horizontal lines, standing for $-\overline k(\omega_p)$ and $\overline k(\omega_p)$, respectively, give the radiation frequencies $\omega_{RR}=4.2\,\omega_0$ ($\lambda_{RR}=575$ nm) and $\omega_{NRR}=5.95\,\omega_0$ ($\lambda_{NRR}=405$ nm). Here $\omega_p \equiv 2\pi c/\lambda_0$ ($\lambda_0=2400$ nm), $v=0.99876\cdot v_g(\omega_0)$ (as derived from Fig. \ref{fig1}), and $k_{NL}=0$ (negligible nonlinear shift). (b) Color level plot of the evolution of the electric field spectrum (ordinary polarization, log scale).} 
\label{fig2}
\end{figure}

As a first example we consider a medium with anomalous GVD ($k''<0$), with SHG occurring in the regime of high negative mismatch $\Delta k=k(2\omega)-2k(\omega)<0$, which results into an effective focusing Kerr nonlinearity, supporting solitary wave propagation \cite{reviewSHG}. These conditions can be realized e.g. in a $\beta-$BaB$_2$O$_4$ (BBO) crystal at carrier wavelength $\lambda_{0}=2\pi c/\omega_0=2400$ nm. By exploiting type I ($o+o\rightarrow e$) SHG in a crystal with orientation angles $\theta=45^o$ and $\phi=90^o$ (quadratic nonlinear coefficients $d_{22}=2.2$ pm/V, and $d_{31}=0.16$ pm/V \cite{nikogosyan}), we obtain from Sellmeier equations a phase mismatch $\Delta k=k_e(2\omega)-2k_o(\omega)=-2.1\cdot10^{-5}m^{-1}$ and a GVD $k''=-0.18$ps$^2/$m. We consider the propagation of an ordinarily polarized hyperbolic secant pulse ${\rm sech}(t/t_0)$ with $t_0=20$ fs duration.  We show typical results obtained for  
%damage threshold so high ?
a soliton number $N=\sqrt{L_d/L_{nl}} \simeq 2$, where $L_d=(t_0)^2/|k''|$ and $L_{nl}=[\omega_0 n_{2I} I/c]^{-1}$ (input peak intensity $7.4$ TW/cm$^2$ in vacuum).
%$L_{nl}=[(2\pi/\lambda_0) n_{2I} I]^{-1}$ are the dispersion and nonlinear length, respectively,
Here $n_{2I}=-\frac{4\pi}{\lambda_0} \frac{\eta_0}{n^2(\omega_0) n(2\omega_0)} \frac{d_{eff}^2}{\Delta k}$ is the effective Kerr nonlinear index due to cascading, $\eta_0$ being the vacuum impedance.
Figure \ref{fig1} shows the time domain evolution of the ordinarily polarized electric field \cite{nota1}. Radiation starts to be emitted at the activation length $z=1.8$ mm, where the maximal pulse compression and spectral broadening are achieved [see Fig. \ref{fig2}(b)]. After this stage soliton fission occurs with the two constituent solitons separating asymptotically. The temporal evolution in log-scale reported in Fig. \ref{fig1}(b) clearly shows that the emitted radiation, which is slower, possesses two distinct branches traveling at different velocities, which turn out to correspond to the RR and the NRR dispersive waves. Indeed the central frequency of these two branches found from the spectral evolution in Fig. \ref{fig2}(b) to be  $\omega_{RR}=4.2\,\omega_0$ ($\lambda_{RR}=575$ nm) and $\omega_{NRR}=5.95\,\omega_0$ ($\lambda_{NRR}=405$ nm), are accurately described by Eq. (\ref{eqRR}). The graphical solution of this equation displayed in Fig. \ref{fig2}(a), shows indeed that such values of $\omega_{RR}$ and $\omega_{NRR}$ are obtained as the intersection of the dispersion curve $D(\omega)$ with the wavenumber of the positive- [$\overline k(\omega_0)$] and negative-frequency [$\overline k(-\omega_0)$] components of the pump pulse, respectively. We emphasize that, while the nonlinear phase shift $k_{NL}$ turns out to negligible in this case, it is of paramount importance to accurately estimate the pump velocity $v$ around the activation length where the radiation is emitted, since the curve $D(\omega)$ is dramatically affected by even small errors in the value of $v$. Here we extract the correct value of $v$ from the time domain evolution, finding $v=0.99876\cdot v_g(\omega_0)$, which correctly describes the pulse velocity at its maximal compression as shown by the black line in Fig. \ref{fig1}(a).
%--fig3
\begin{figure}[tb]
\includegraphics[width=8cm]{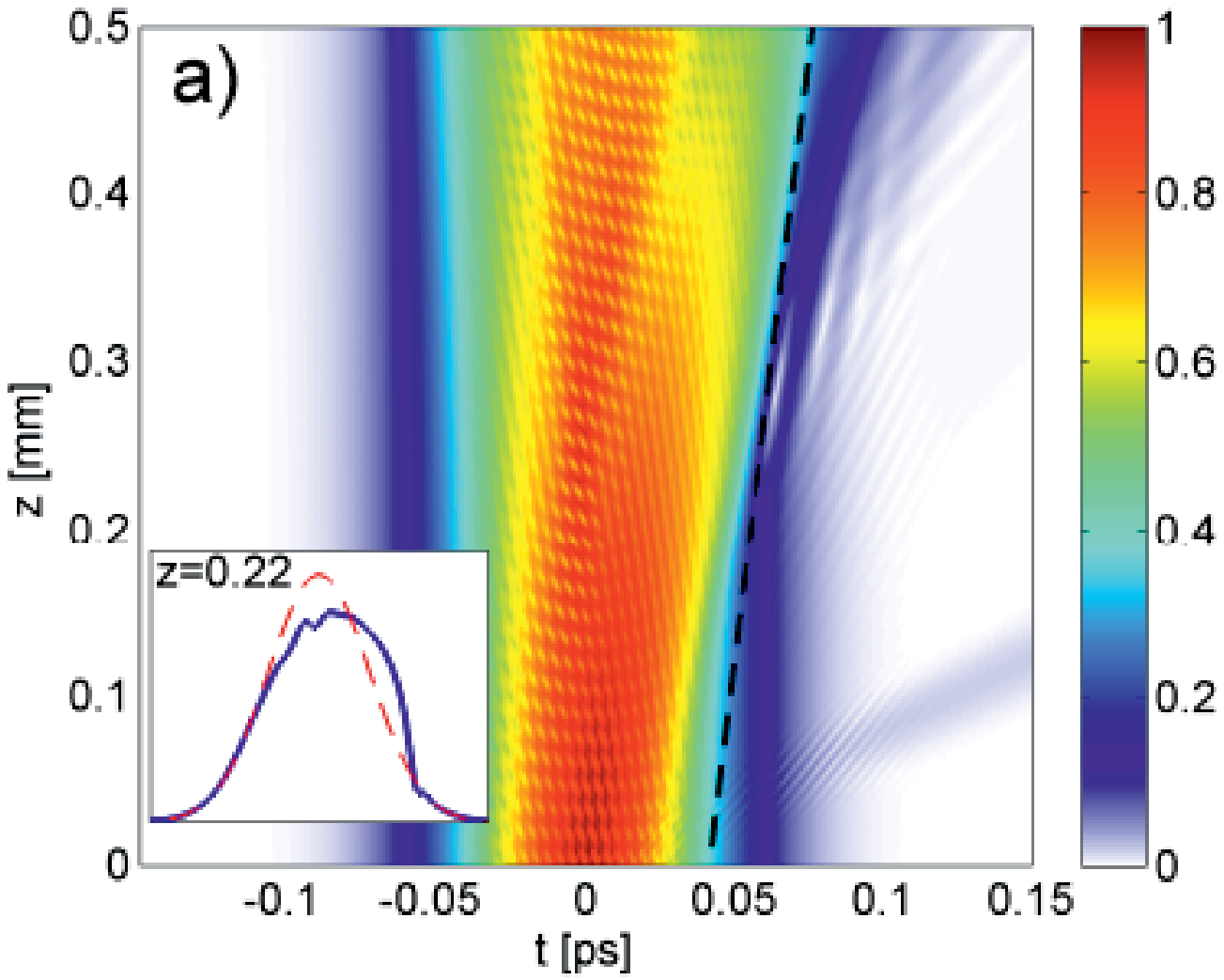}
\includegraphics[width=8cm]{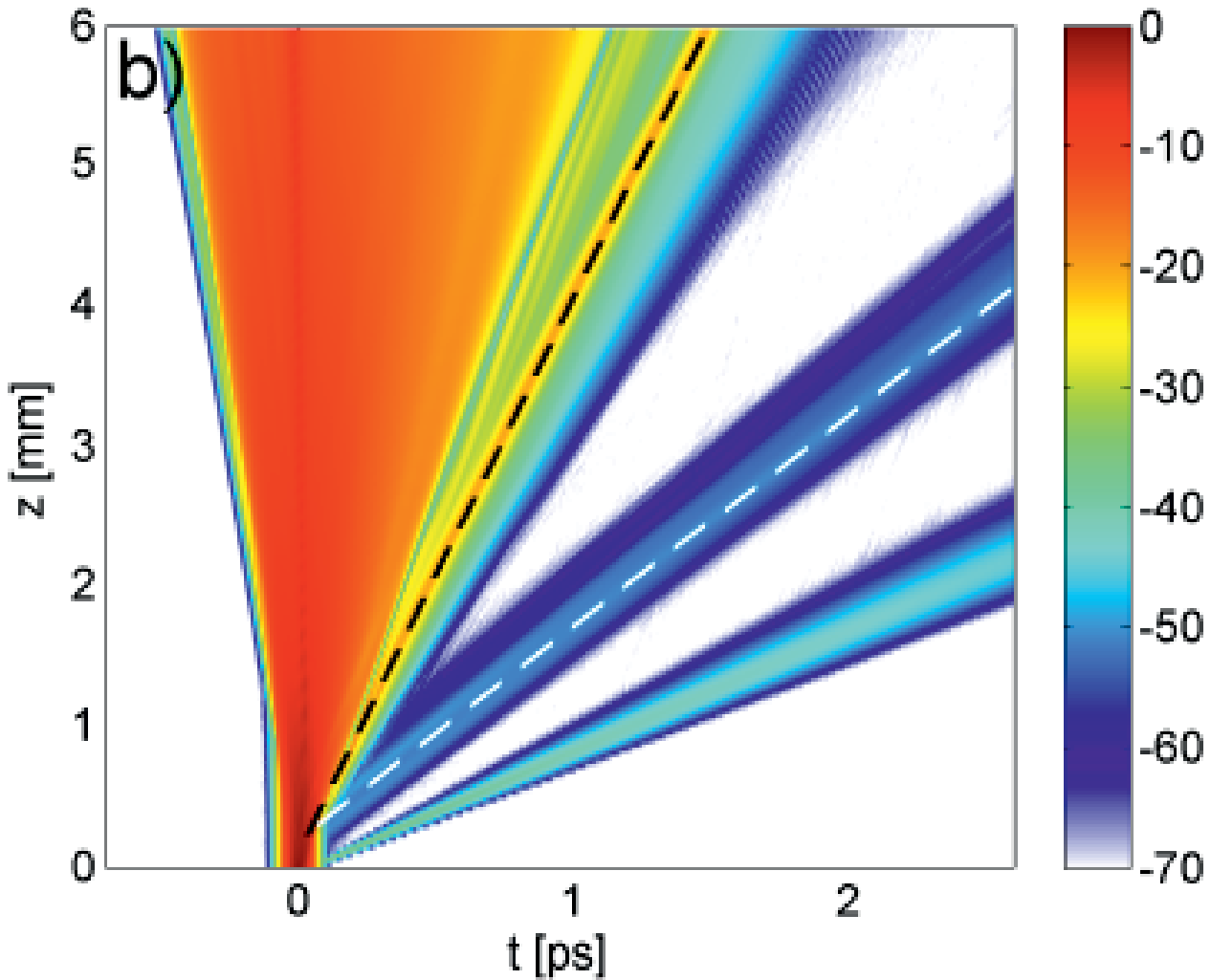}
\caption{(Color online) Color level plot of time domain evolution of the electric field \cite{nota1} (ordinary polarization): (a) linear scale, early stage ($z \le 0.5$ mm). Inset: snapshot at the point of shock formation compared with the input. The black line stands for the shock velocity $v=0.993 \cdot v_g(\omega_0)$; (b) log (dB) scale, long range evolution ($z \le 6$ mm).
The dashed black and dashed white correspond to the RR and the NRR, respectively. 
%(see Fig. \ref{fig4}).
} 
\label{fig3}
\end{figure}
%-- fig4
\begin{figure}[t]
\includegraphics[width=8cm]{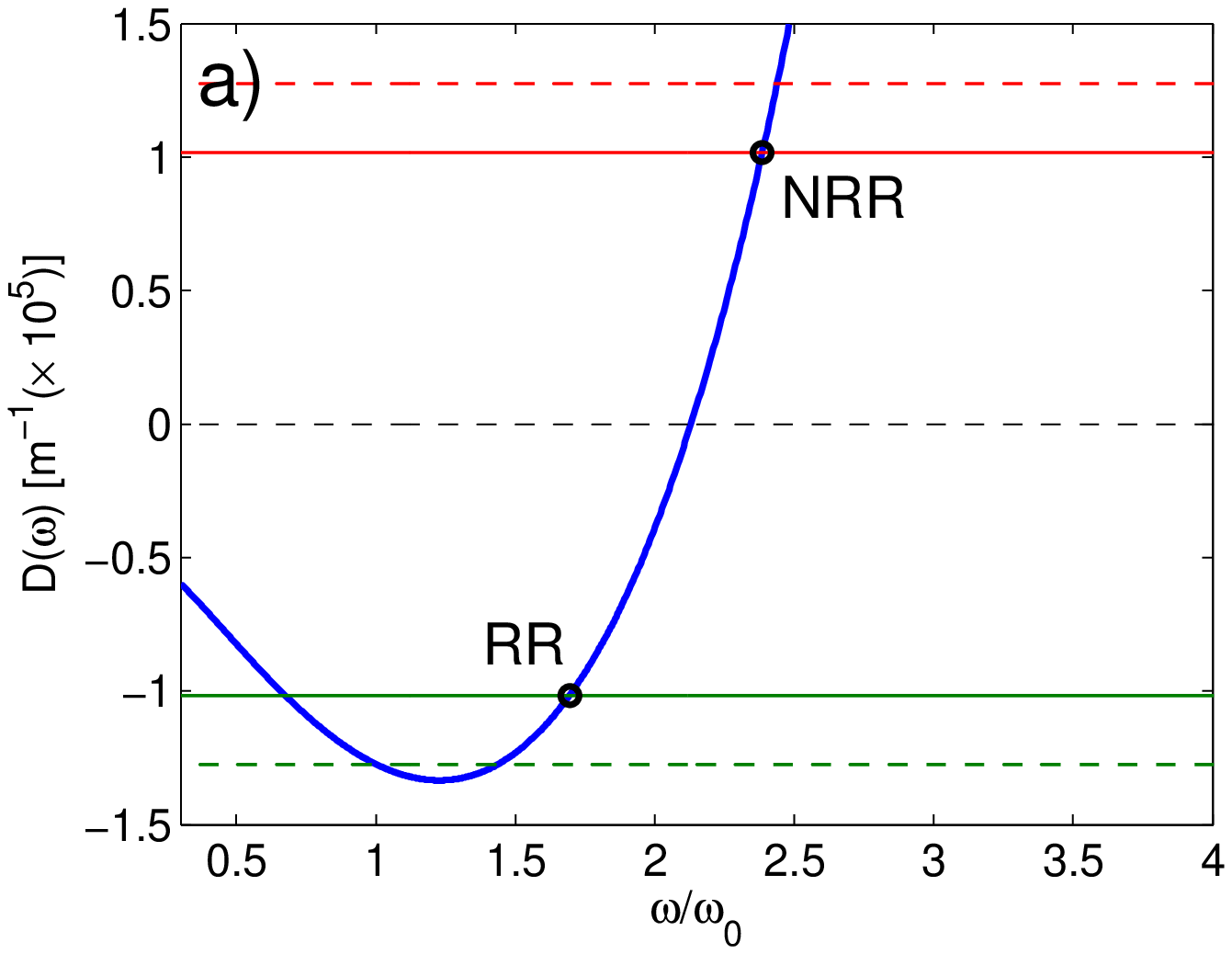}
\includegraphics[width=8cm]{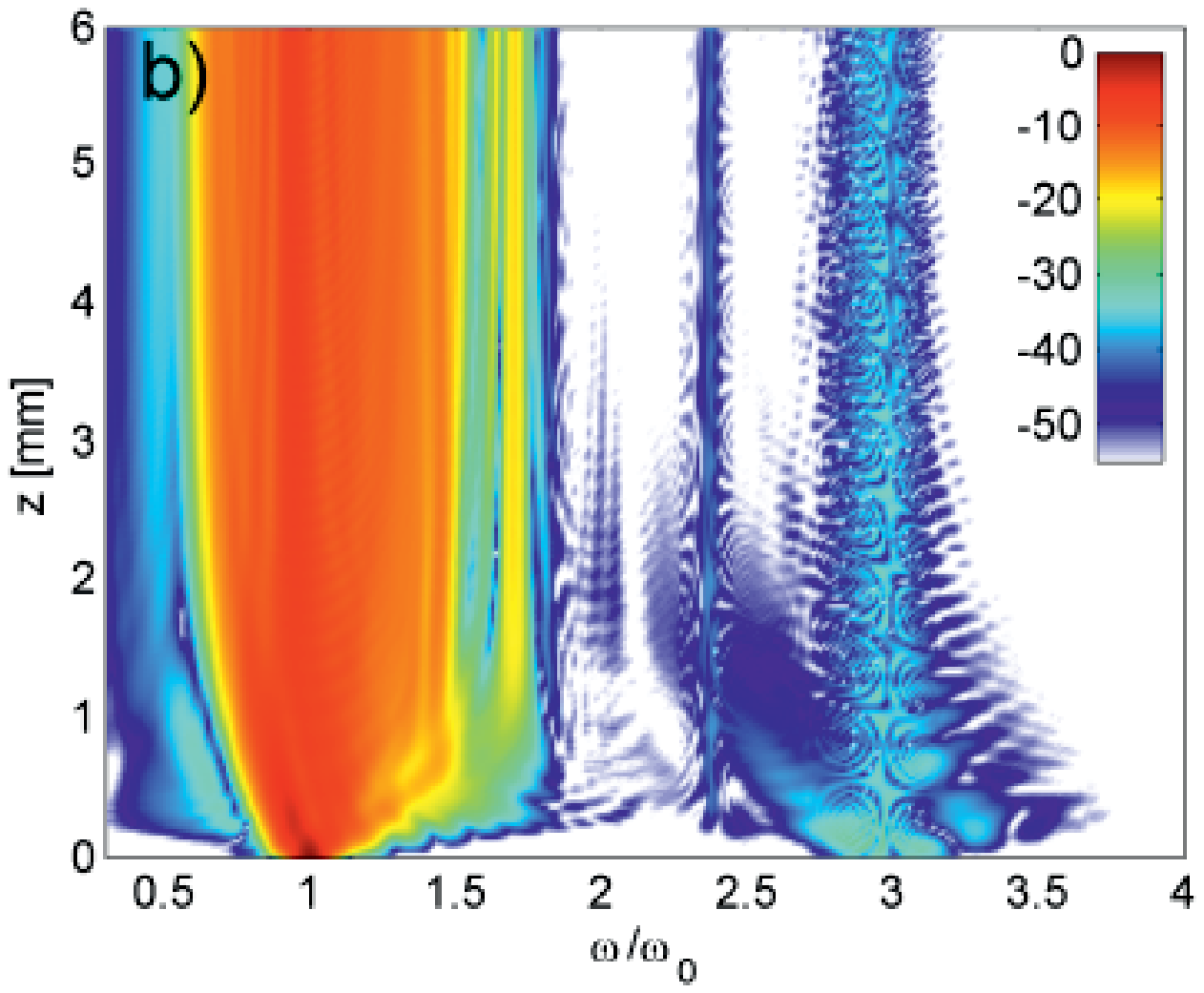}
\caption{(Color online) As in Fig. \ref{fig2} for the case of wave-breaking in GaSe described in Fig. \ref{fig3}. In (a) the graphical solutions yielding $\omega_{RR}=1.7\,\omega_0$ ($\lambda_{RR}=1410$ nm) and $\omega_{NRR}=2.37\,\omega_0$ ($\lambda_{NRR}=1010$ nm) are obtained with the full expression of $\pm \overline k(\omega_p)$ (solid horizontal lines), while for comparison the dashed lines stand for the corresponding quantities calculated with $k_{NL}=0$.} 
\label{fig4}
\end{figure}

In the second case we consider the opposite sign of dispersion, namely normal GVD, yet with the same sign of mismatch. 
Under these conditions, the cascading nonlinearity does not compensate for GVD-induced temporal broadening but rather enforces it. 
We have carefully chosen the operating conditions to work in a regime where the nonlinearity initially dominates over the dispersion (i.e. weakly dispersing regime), which is characteristic of the formation of DSW \cite{Rothenberg89,Hoefer06,Wan07,Conforti12}. In this regime, we find viable conditions for the observation of NRR  e.g. in Gallium Selenide (GaSe), which is characterized by a large nonlinear coefficient $d_{22}=54$ pm/V \cite{nikogosyan}. By operating at central wavelength $\lambda_{0}=2400$ nm with type I ($o+o\rightarrow e$) SHG in a crystal oriented at angles $\theta=32^o$ and $\phi=90^o$, we find from the Sellmeier formulas, a phase mismatch $\Delta k=-3.5\cdot10^{-5}m^{-1}$, and a GVD $k''=0.32$~ps$^2/$m.
We show in Figs. \ref{fig3}-\ref{fig4}  the outcome of our simulations obtained from a ordinarily polarized input pulse with $50$ fs duration (FWHMI), gaussian shape, and input peak intensity $500$ GW$/$cm$^2$ (in vacuum). In this case $L_d$ is several orders of magnitude larger than $L_{nl}$, and the dynamics is essentially dominated by the nonlinearity. The latter is responsible for the pulse temporal broadening and steepening shown in Fig. \ref{fig3}(a). In particular steepening is found to occur on the trailing edge until a gradient catastrophe leads to the formation of a shock wave (maximal steep front) at $z = z_s \simeq 0.22$ mm [see inset in Fig. \ref{fig3}(a)]. Here the dynamics is essentially different from Kerr media where two symmetric shocks are formed over the leading and trailing edges \cite{Rothenberg89}, as also confirmed recently with reference to spatial dynamics \cite{Hoefer06,Wan07}. The reason is that, in SHG the repeated up- ($\omega+\omega = 2\omega$) and down-conversion ($2\omega-\omega = \omega$) gives rise, not only to the well known effective Kerr nonlinearity, but also to a leading-order steepening term \cite{Ilday04}, owing to the group-velocity mismatch, which induces the shock to be asymmetric \cite{Conforti12}. Whenever the group velocity at fundamental frequency is sufficiently larger than that at second-harmonic, this term dominates and leads to shock formation on the trailing edge \cite{Ilday04,Conforti12}. Once formed, the shock front travels with a characteristic velocity which we estimate numerically to be $v=0.993\cdot v_g(\omega_0)$, while it develops fast oscillations due to the GVD [these occur, in this case, on a small scale due to the absence of a pulse background \cite{Conforti12} and hence are not visible in Fig. \ref{fig3}(a)]. The shock formation is accompanied by an abrupt spectral broadening [see Fig. \ref{fig4}(b)] and the consequent emission of radiation. The latter is emitted along two branches, as shown by the temporal evolution in log-scale displayed in Fig. \ref{fig3}(b). The different speeds of these branches arise from their different frequencies, which are found from the spectrum in Fig. \ref{fig4}(b) to be $\omega_{RR}=1.7\,\omega_0$ ($\lambda_{RR}=1410$ nm) and $\omega_{NRR}=3.7\,\omega_0$ ($\lambda_{NRR}=1010$ nm). Also in this case we denoted such frequencies as RR and NRR since they agree perfectly well with the values obtained by the graphical solution of Eq. (\ref{eqRR}), illustrated in Fig. \ref{fig4}(a). Again, it is crucial to have an accurate estimate of the pump velocity $v$ in the proximity of the radiation emission distance ($z \simeq z_s$).
However, in this case the larger nonlinear coefficient $d_{eff}$ results in a non-negligible $k_{NL}$. Indeed we predict the correct values of $\omega_{RR}$ and 
$\omega_{NRR}$ by evaluating $k_{NL}$ as arising from from local self-phase modulation, i.e. $k_{NL}=\omega_0 n_{2I} I(z_s,t_s)/c$, being $(z_s,t_s)$ the location of the shock. We point out that, unlike the previous example, here the two RR and NRR frequencies are red- and blue-shifted, respectively, with respect to the second-harmonic frequency, whereas Fig. \ref{fig4}(b) shows also a weak component at $3\omega_0$ that corresponds to the slowest wave in Fig. \ref{fig3}b, i.e. non-phase-matched third harmonic generation from the cascaded process $\omega_0+2\omega_0\ = 3\omega_0$.

Finally, we point out that $\chi^{(3)}$ nonlinearities can compete with quadratic ones due to the high intensities (especially in BBO case) involved. However, both BBO and GaSe exhibit a focusing Kerr nonlinear index, which simply results in a lowered intensity threshold for radiation emission without any significant change to the dynamics illustrated above.
This is confirmed by additional simulations (not reported) where we account also for the intrinsic Kerr nonlinearity.
 
\emph{Conclusions}. In summary, we have demonstrated the generality of the NRR phenomenon, by predicting that it can be observed in quadratic media,
under different scenarios that involve pumping either with soliton-like pulses or, in the opposite regime, when pulses undergo wave-breaking.
This represents a substantial step forward towards the understanding and management of ultrafast cascading nonlinearities for producing 
broadband emission (supercontinuum) in standard crystals.

Funding from MIUR (grant PRIN  2009P3K72Z) and EPSRC (grant EP/.J00443X/1) is gratefully acknowledged.
%\pagebreak

%------------------

\begin{thebibliography}{}
\bibitem{nota1}
We display the inverse Fourier transform of the positive frequency components of the spectrum. This allows us to directly visualize the envelope, eliminating the fast oscillations of the carrier while retaining the full broadband features of the spectrum.
\end{thebibliography}
\end{document}